# SPECIFIC ASPECTS OF INTELLECTUAL PROPERTY MANAGEMENT IN THE KNOWLEDGE-BASED ECONOMY

Aurel Mihail Țîțu [1], Alina Bianca Pop [2], Camelia Oprean-Stan [3], Sebastian Emanuel Stan [4]

[1] Lucian Blaga University of Sibiu, 10, Victoriei Street, Sibiu, România; The Academy of Romanian Scientists, 54, Splaiul Independenței, Sector 5, Bucharest, Romania, mihail.titu@ulbsibiu.ro
[2] SC TECHNOCAD SA, 72, Vasile Alecsandri Street, Baia Mare, Romania, bianca.bontiu@gmail.com
[3] Lucian Blaga University of Sibiu, 10, Victoriei Street, Sibiu, România, camelia.oprean@ulbsibiu.ro
[3] Land Forces Academy Nicolae Balcescu, Sibiu, Romania, sebastian.stan@armyacademy.ro

ABSTRACT: This paper addresses the issue of intellectual property management in the knowledge-based economy. The starting point in carrying out the study is the presentation of some concepts regarding in the first phase, the intellectual capital. Arguments are made that the knowledge-based economy is a challenge for the current century. The subject of intellectual property is approached through the prism of a topical concept operationalized in the current global economic context. The main institutions that are directly related to this concept are mentioned. The topic of patents related to WOS indexed scientific papers is also debated, along with a series of statistics and studies on the state of patent protection worldwide in the top fields. The last part of the paper contains the conclusions and own points of view on the debated topic.

## 1. INTRODUCTION TO THE APPROACH FIELD

Intellectual capital involves a package of activities such as collecting, coding and disseminating information, as well as acquiring new skills based on training and development, including in the process of (re) designing a business.

Intellectual include the knowledge that is held by all people in an organization. In this way, these people give the organization a competitive advantage in the market.

Nowadays, this concern is essential in both developed and developing countries. The argument put forward in this regard is that solutions are being sought for economic growth, development and competitiveness, but also for the purpose of increasing employment welfare.

The protection of intellectual property is imperative because it generates opportunities to create new technologies. The efficient use of these new technologies ensures a strategic advantage in terms of status on the economic and political map of the world.

Until now, the importance of intellectual property has gradually increased considerably. Today, the components of intellectual property are elements of a series of major international treaties but also topics of debate on various current issues, traditional knowledge, innovation, technology transfer, scientific research, ecology, biodiversity, biotechnology, internet, entertainment industries, mass media etc. [1].

When it comes to new economic trends, some of the commonly used concepts are organization, economics and knowledge-based management. At the heart of this is the 21st century knowledge revolution, because it brought "knowledge" to the fore. This concept is very important to ensure that it ensures the functionality and efficiency of organizations. The knowledge-based society is based on the consideration that rigorous learning is necessary through a process of transforming information into knowledge (which can be new products, new technology, etc.).

The transition to the knowledge-based economies of nations has led to understanding of the significant role that knowledge plays in economic growth. Thus, managers seek to capitalize on the knowledge incorporated by employees and information system designers try to incorporate knowledge into technologies based on databases and logic programs.

## 2. INTELLECTUAL CAPITAL IN THE KNOWLEDGE-BASED ECONOMY

The knowledge-based organization is built on a group of people who make up its own structure and work together to achieve its goals [2]. This type of knowledge-based organization highlights new and different perspectives on how to develop and apply management practices.

2.1 The knowledge-based economy - a challenge of the current century

Knowledge is an idealized concept based on the fact that information is an accumulation of data organized quite easily and being easy to assimilate, but which need knowledge in order to be capitalized. Knowledge can be: tacit or explicit, factual or inferential.

The knowledge-based economy consist on the production, distribution and knowledge use in order to create goods and services that serve sustainable development.

In the recent global economy, the share of intellectual capital has increased in direct proportion to the number of knowledge-based industries. "

The OECD's definition of the knowledge-based economy is limited to investments in knowledge that represent research and development expenditures.

Today, organizations are adapting their development strategies, and in this way they acquire a new vision.

The distinction between the classical organization, authority structures and new knowledge-based organizations is important and can cause problems with long-term benefit management.

In the new current and future context, it is essential to understand the meaning of resources. Knowing the resources is an advantage that can avoid the dangers of the prognosis. This can lead to improvements and efficiency. Through their specific culture, knowledge-based organizations include an increasing conceptual value in society as a factor of influence as well as a source of competition.

It should be noted that the relationship between the knowledge-based economy and the construction and the organizational function of the knowledge-based cannot be established without knowledge-based management [2, 4].

The concept of knowledge management is treated by Brătianu [5], along with strategic thinking [6].

Knowledge of the human capital is at the organizational level, and the needs and preferences of the consumer are specified in the products, processes, capabilities, and systems that produce the system's capital plans.

The value of knowledge can greatly increase the value of intangible costs. However, intellectual property rights are often not given proper consideration. But in knowledge-based organizations, decision-making is very important [2].

The speed with which new brands, products, models and models enter the market should not be underestimated, because innovation and innovation in today's world are critical. The phrase "use of knowledge and information technology" refers to one of the important elements of the driving force that has brought significant changes in the work of different companies. These changes, and for many consumers, are a constant challenge for companies that want to focus on the resources that have been denied. These characteristics have not been seen in the past, and today the technological authority has become an important factor in ensuring competitive advantage [7, 8].

In the context of the modern economy, the importance of value creation is the convergence of artificial intelligence or knowledge strategies, so that values and benefits cannot be easily distorted.

Recognition of information and values (name, trust in customers) are effective ways to change. In this way, the old model of defining accounting, management and finance does not reflect the current dream. In this way, new ways of reporting the correct powers of the mind appear. This is a competitive advantage that many organizations are not aware of.

Vulnerable assets, which include intelligence, knowledge, intellectual property and experience, are opportunities for future development and reflection. The price must be calculated to increase the market value of the organization (Figure 1).

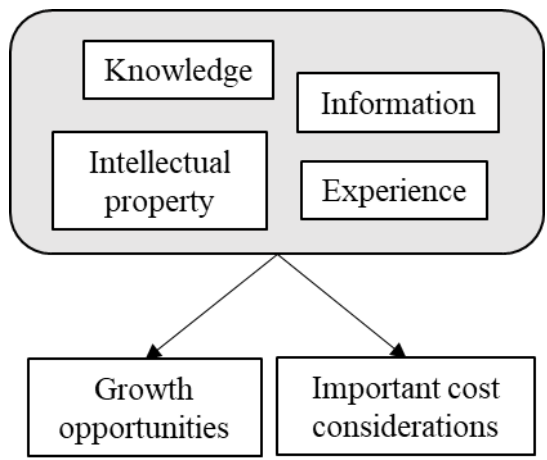

**Figure 1.** importance of intangible assets

Specifically, the primary market and the asset that cannot be a competitive advantage can be used to differentiate between market value and group book value over time. Reduce the use of traditional financial information [8, 9].

Development of different types of communities Research on quarantine development is correlated with community achievement with high results, investing in their innovations [10].

In recent decades, the knowledge-based economy has been an element of innovation in developing countries. This aspect is completely different from the industrial economy or its post-industrial version.

A transition to a knowledge-based economy, knowledge-based management is essential for building and implementing knowledge-based organizations.

In the case of organizations that have a successful portfolio to attract small and medium-sized enterprises and foreign investors, especially those without internal sources of funding, patents are considered a means of attracting and securing investment (Figure 2).

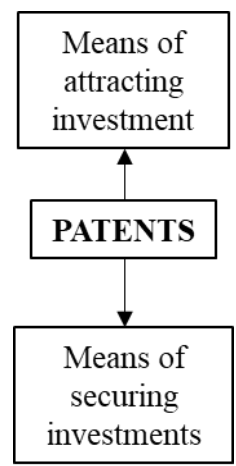

**Figure 2.** The importance of patents

Investors are attracted to the fact that companies have a technical advantage over their competitors protected by patent law if they have a strict portfolio of intellectual property rights [11,12].

Innovation is on the rise because some organizations are better than others and can compete on global innovation issues.

However, the main task of the business is to preserve this innovative mindset and protect the value of the creative process by only taking into account financial assets and related financial resources.

Therefore, even if it is already more aggressive, steps must be taken to support innovation or continue existing innovation. These measures need to support innovation, education, finance, tax creativity, and within each company.

In reality, there are few rules and information cannot be effectively used to reflect the value of recognized intangible assets or intangible assets. The recognition of intangible assets is actively discussed in traditional statistical literature.

2.2 A pragmatic approach to the concept of intellectual capital

One way to define intellectual capital would be to sum up all the knowledge held by all persons working in an economic entity and at the same time giving that entity a competitive advantage in the market.

Intellectual capital must be assessed both in terms of its impact at the functional level and in terms of the impact at the global organizational level. Although, erroneously, intellectual capital and especially the human capital component, are perceived as a source of costs, it must be considered as a strategic investment, analysing all the components shown in Figure 3.

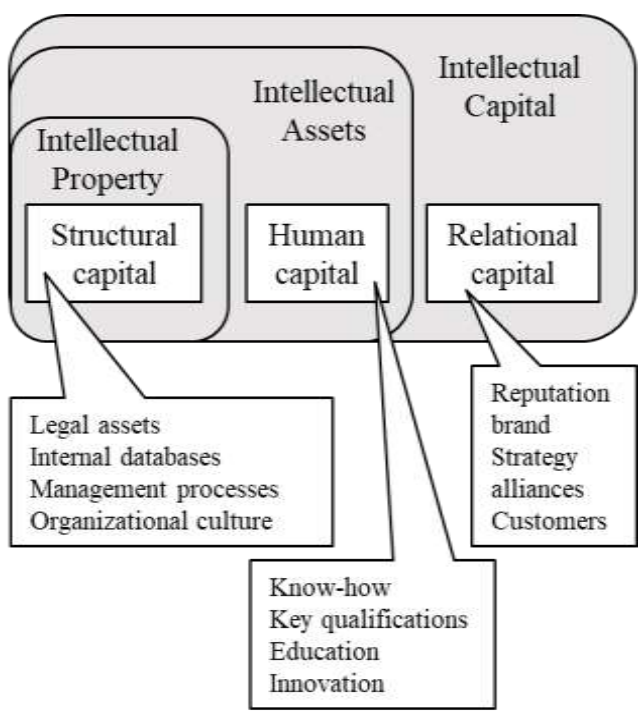

**Figure 3.** The structure of intellectual capital

The concepts of intellectual property and intellectual capital must be closely linked. Japan, the United States and Europe were the first to draw attention to these concepts.

In Europe, Karl Erik Sveiby (1997) [13, 14] contributed to the development of intellectual capital. Intellectual property is a legal form that allows the owner to control certain intangible assets, such as ideas or expressions.

According to the Romanian Copyright Office (ORDA, 2017) [15], "Intellectual property, abbreviated IP, considers the creations of the mind: inventions, literary works, works of art, symbols, names, images and design used in commercial activities. Intellectual property has the ability to control and reward its use, so as to encourage innovation and creativity for the benefit of humanity."

In terms of organizational performance, the economic-financial part has always been an important factor in decision-making. Correlating with the principle of quality management according to which decisions must be made based on facts, a careful analysis of financial reports, accounting documents is imperative for stakeholders to make the best decisions. However, in the age of knowledge, when intellectual capital is an important part of the value of a product, it is not well enough represented in classical financial documents.

An analysis of the literature can conclude that organizational results are not the result of macroeconomic or financial policies, but are an undesirable result of technological progress, innovation, the quality of human, structural or relative capital and investment in knowledge - education, research and development.

## 3. INTELLECTUAL PROPERTY - UPDATE CONCEPT OPERATIONALIZED IN THE CURRENT GLOBAL ECONOMIC CONTEXT

3.1 Defining the concept of Intellectual Property and applicability

Intellectual property is a key factor in cultural, social and economic development. Its components are industrial property and copyright and related rights (Figure 4).

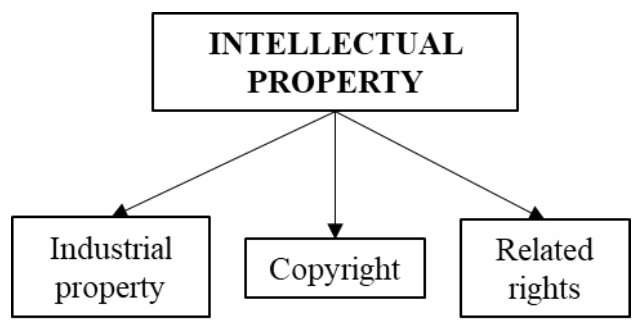

**Figure 4.** Components of intellectual property

In this way, it can be mentioned that the protection of intellectual property rights plays a significant role. Its main result is the protection of the product of human intelligence, as well as the benefit of consumers to be able to use this product [16].

Bill Gates said that "Intellectual property has the term of validity of a banana." Thus, the protection of intellectual property is absolutely necessary, due to the opportunity to create new technologies and use them efficiently, which offers a strategic advantage. The argument that supports this statement is the position of the statute on the economic and political map of the world that is determined. Another argument is the progress and prosperity of the country, which is influenced by the creativity of its citizens, both technically and culturally. In terms of legal protection of new creations, they encourage investment and lead to new innovations. Another argument is the promotion and protection of intellectual property that stimulates economic growth and, at the same time, leads to the creation of new jobs and creates new branches of activity and, implicitly, the improvement of the quality of life.

When it comes to the knowledge society, intellectual property has a key role to play in protecting knowledge against alienation and illegal use. In this sense, incentives are offered to innovators to generate new ideas and concepts.

Intellectual property is an indispensable attribute of the market economy and, at the same time, a key engine of the world economy in globalized trade.

Major contributors to the economy are represented by industrial sectors based on the protection of intellectual property rights. Thus, a clear, robust and efficient intellectual property rights regime is a fundamental and imperative condition for attracting foreign investment directly, based on promoting research and development and technology transfer.

Intellectual property is an indeterminate attribute of the market economy and, at the same time, a key factor of the world economy in the conditions of globalized trade.

3.2   Institutions that are directly related to this concept

In 1893, the United Nations Department of Intellectual Property Protection, BIRPI, was founded. Then, in 1967, the World Intellectual Property Organization (WIPO) was born. In 1974, SIPO became a specialized agency of the United Nations. The head office is located in Geneva. The total number of member states was 183 and the number of international agreements was 23. Its purpose was to promote creative activities, including the promotion and protection of intellectual property.

In our country, the Office of Inventions and Trademarks (OSIM) protects intellectual property related to industrial property rights, and ORDA protects copyright and related rights.

ORDA, as a specialized agency that depends on the Law No. Government, for rights related to copyright, August 1996. A single agency across Romania that regulates registration in the national register, as well as supervision, approval, arbitration, scientific and technical conclusions in the field of copyright. Rights as well as areas of rights.

The International Intellectual Property Organization (WIPO) was established in 1967. The organization argues that intellectual property is "the right to literature, art and science, to the definition of entertainment and entertainment type, audio and video processing, innovations in all aspects of human activity, scientific discoveries and much more."It represents the industry of production, trading, trade and services. "" Trade names, trade names, protection against unfair competition, intellectual property rights, science, literature, visual arts in other fields"

In 1974, the organization was established by the United Nations, based in Geneva, and brought together 183 member states and 23 international treaties to encourage action and promote and protect intellectual property.

The Counterfeiting Agreement (ACTA), introduced by the United States, the European Union, Switzerland and Japan in 2002, seeks to counter the growing world trade in counterfeit products and copyrighted works. Without the author's permission.

These discussions were attended by Australia, South Korea, New Zealand, Mexico, Jordan, Morocco, Singapore, the United Arab Emirates, Canada, Jordan, Morocco, Singapore, the United Arab Emirates and Canada.

However, this agreement was not submitted before the public hearing before 2011, the main objective of which:

- Support for economic growth in all sectors;

- Counterfeiting goods and eliminating risks for consumers and reducing organized crime;

- to promote the effective practice of intellectual property, taking into account the differences between their systems and practices at national level;

- ensure a balance between the rights and interests of the owner of intellectual property rights, service providers and consumers;

- Ensure cooperation between service providers and rights holders in a digital environment.

This agreement has long been criticized, both for the secrecy of the negotiations and for the fact that it could lead to a cult of surveillance and suspicion.

In July 2012, the European Parliament finally rejected ACTA, depriving the EU of the opportunity to ratify this controversial international agreement, which critics say threatens free people, especially those connected to the internet.

The European Union Intellectual Property Office (EUIPO) retains its rights in trademarks, industrial designs registered in the European Union at the European Astronomical Observatory and is a database on orphan works since 2012.

The European Patent Office (EPO) is one of the two agencies of the European Patent Office (EPO), the other being the Board of Directors.

The EPO acts as the agency's director, but the council functions as a regulatory authority and to some extent as a legislator.

The true legislative power to revise the European Patent Agreement belongs to the Contracting States even when they meet at the Conference of the Contracting States.

The patents granted by the EPO apply to European countries where applicants wish to patent their inventions.

Because EPO patent applications are about three times more frequent in most European countries than national applications, EPO patents have a high cost of inventory, which requires protection in many European markets.

However, centralizing and examining the application has also been a central legal challenge: under the European Patent Convention (EPC), all third parties can file a counterclaim to apply for a patent in all designated states within nine months of upon receipt of the show. In the state [18].

The United States Patent Office (USPTO) is an agency of the United States Department of Commerce that grants patents to inventors and businesses for the invention and trademark for the identification of products and intellectual property.

The USPTO was founded in Alexandria, Virginia, in 2005, near the city of Crystal, after moving to the neighboring city of Arlington, Virginia.

The patent offices and the main undercover agent who remained at the southern end of Crystal City completed the transition to Randolph Square, the brand's new building, in the village of Shirlington on April 27, 2009.

The Japanese Patent Office is a Japanese government agency responsible for industrial property rights located at the Ministry of Economic Affairs. The Japanese patent office is located in Kasumigasek, Chioda and Tokyo and is one of the largest patent offices in the world. The role of the Japan Patent Office is to promote the growth of the Japanese business sector by managing patents, data models, projects and trademarks. (Copyright is administered by the Cultural Institute.)

The first patent law in Japan was enacted in 1871, but was repealed the following year. Today, April 18, 1885, when the Japanese Patent Law was founded and the Patent Office of Japan, when the Patent Law was adopted. In 1899, Japan became a party to the Paris Convention for the Protection of Industrial Property. Takahashi Korekiyo was the first CEO of JPO.

3.3   Presentation of the main common forms of Intellectual Property

Figure 5 summarizes the elements of intellectual property.

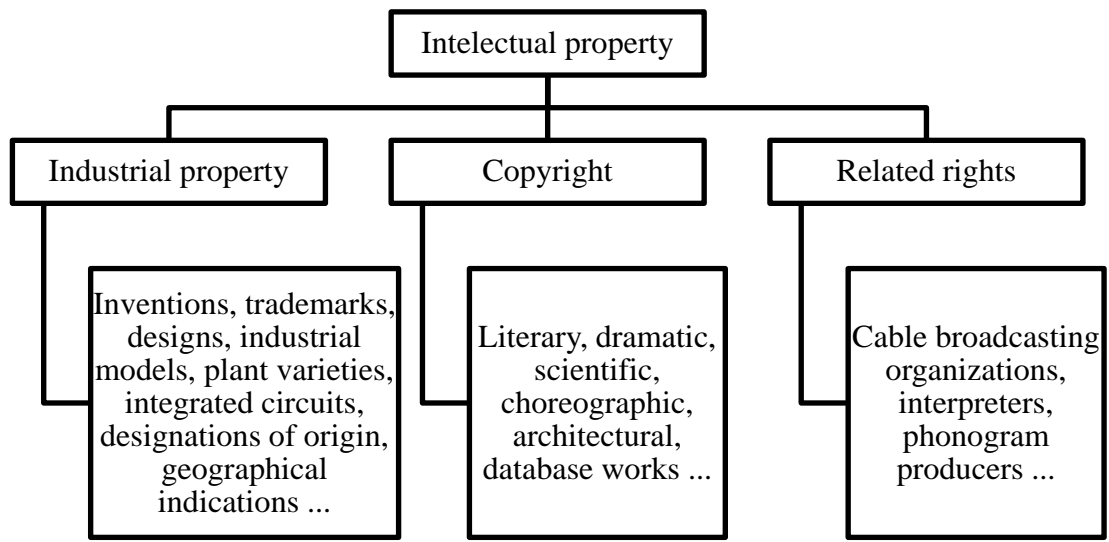

**Figure 5.** The components of intellectual property

3.4   Patent versus WOS Indexed Scientific Papers

The main purpose of scientific research is to acquire knowledge. These, being capitalized later, offer multiple benefits to humanity. One of the most important forms of knowledge enhancement is scientific publications, which are an objective and relevant reflection of the most important results of research activities. Scientific information reflected in publications is usually stored in books (especially treatises and monographs) that integrate articles, theses, research reports, memoirs of scientific conferences (proceedings), doctoral theses, patents, grants research (related scientific reports), etc., all of which are printed on paper and / or electronically.

The validation and capitalization of the obtained results, reflected in different publications, is done by disseminating them in the scientific community.

Ideas can remain in a person's intellect, can be published in specialized journals or, when they meet certain conditions, can be protected by methods specific to the protection of intellectual property, in particular industrial property such as: patent, utility model, topography of TPS semiconductor products , industrial design, brand, etc.

A patent is a document that grants a set of exclusive rights by a country to an inventor or a proxy chosen by the inventor for a fixed period of time in exchange for disclosing the invention. The procedure for granting the patent, the conditions to be met and the period for which the exclusive rights are granted vary widely from country to country in accordance with national law and international agreements. A patent application must include one or more claims that define the invention as something new, something that requires creativity and can be applied industrially. In most countries, computer programs and business methods, both accepted in the United States, are excluded from patenting. The exclusive rights granted by the patent in most countries are to prevent or exclude others from manufacturing, using, selling, and importing the product under the patent.

In order to obtain the patent in Romania, an application for a patent is filed with OSIM. The application will contain, in addition to a standard form, a documentation describing the object of the application.

The patent may provide rights in several countries around the world, but it is necessary to complete the patent formalities in each country. The PCT, the Patent Cooperation Treaty, is an international treaty on patent law established in 1970, which provides for a single procedure for the protection of the invention in each country party to the treaty. The "PCT application" procedure ensures that the inventor's priority is recognized in the signatory countries by allowing him to file a patent application in each country within a period of up to 30 months. Otherwise, the international patent cannot be made later than one year after the filing of the first patent application.

In most cases, the only intangible assets recognized in financial-accounting reporting are patents and trademarks. But, although in the case of other elements that are part of what is generically called intellectual capital, it is difficult to make an assignment of value, they must be considered and considered in order to understand the process of creating value.

Patents give the owners an exclusive right over an invention, whether it is a product or a process, which aims to bring a new technical solution to solve a problem.

Patents are a useful indicator for measuring research results. In particular, the number of patents is used to determine the innovative capacity of a region, the level of knowledge dissemination and the level of internationalization of innovative activities (Figure 6).

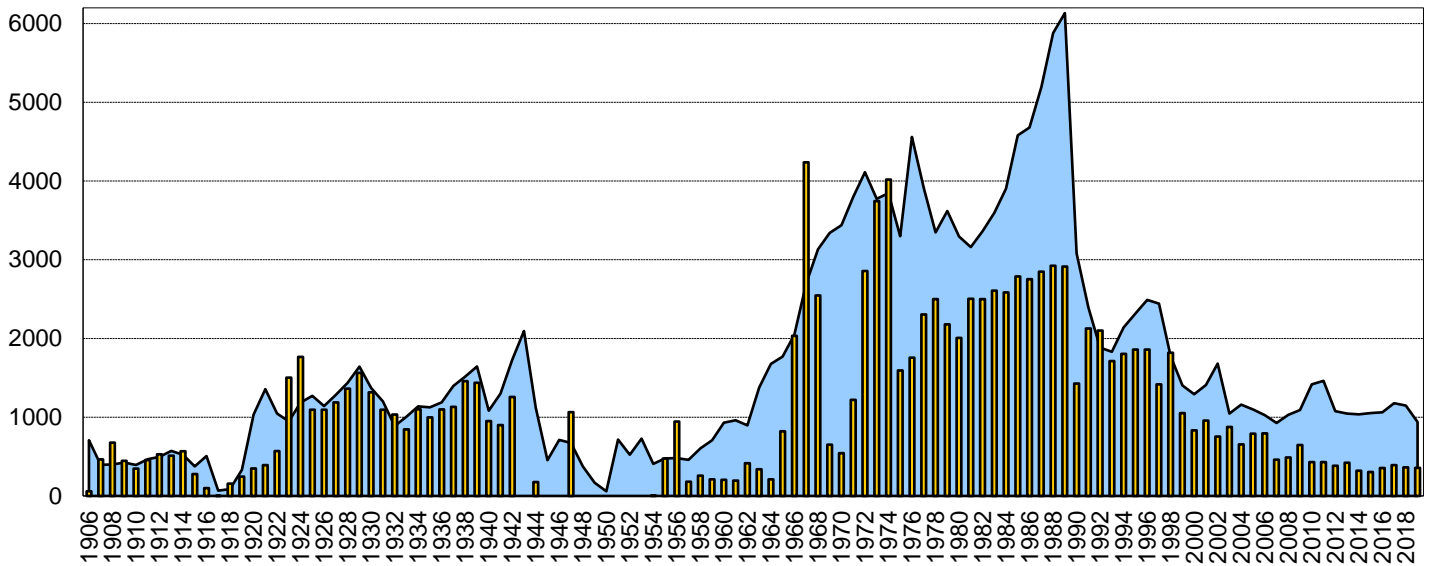

**Figure 6.** Filed patent applications and patents granted by the OSIM in the period 1906 - 2019 [19]

Patent applications are the first step in the patent application process which indicates the potential interest of innovative organizations around the world in the European technology market. Analysing the situation at European level, from the following figure it can be seen that the number of patent applications filed at European level increased in 2018 by 4.6%, reaching a new level of over 174,317 (Figure 7).

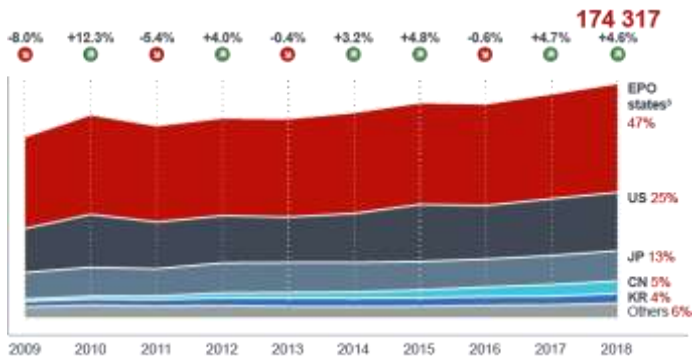

**Figure 7.** The geographic origin of the European patent applications filed with the EPO [20]

Regarding the origin of European patent applications, it can be seen that they came from EPO Member States, followed by the USA, Japan, China and South Korea. The number of patent applications in Europe was identified with those of the previous year, with significant differences between countries. The European growth champions were Belgium and Italy. More modest increases came from Austria, Spain, Switzerland and the United Kingdom. The highest number of patent applications in Europe was in Germany, while France, the Netherlands and many of the Nordic countries filed fewer applications. Overall, China (and to a lesser extent South Korea) has been the main driver of growing EPO demand. US applications have dropped significantly (a cause may be changes to US patent law), while Japanese organizations have filed fewer applications (Figure 8).

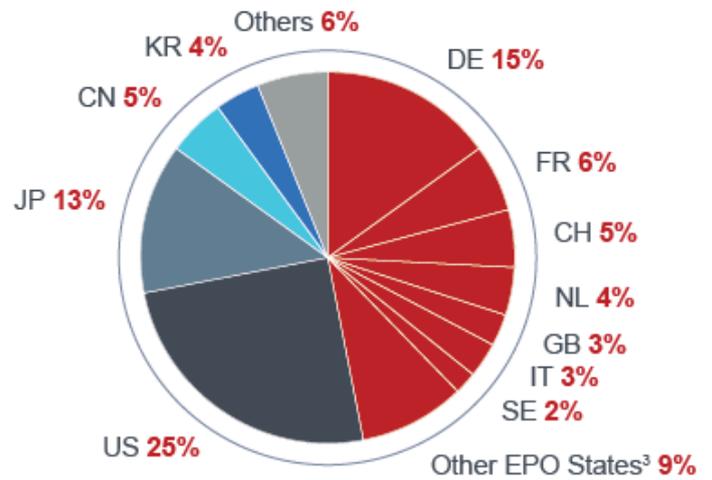

**Figure 8.** The geographic origin of the European patent applications determined by the country of residence of the first applicant listed on the application form [21]

2018 was a representative year in terms of the number of patent applications in the world. China continues to be the world's largest patent applicant. In 2018 alone, Chinese inventors filed 154, 2002 applications for protection of inventions, placing China on the first step of the podium. China is followed by the USA with 597141 and Japan with 313 567 applications for protection of inventions (figure 9).

26

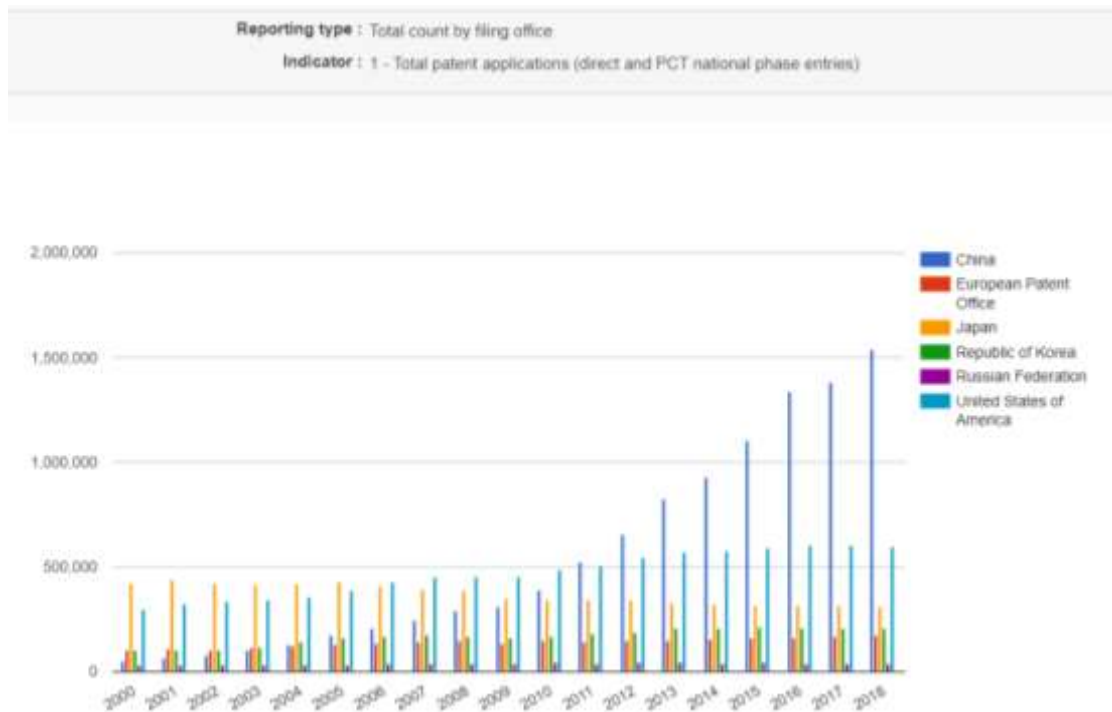

**Figure 9.** Total patent application [22]

Analysing the countries from which most patent applications have come is necessary to draw a parallel with their level of economic development.

Most countries with obvious inventive step are part of the G20, the group of the most advanced economies in the world.

The direct relationship between a strong economy and a modern innovation system no longer needs to be demonstrated, it must be implemented through policies, education and information.

The world's leading economic powers are also the most active in terms of innovation. Organizations and universities in China, the United States and Japan have filed almost half of the total number of patent applications.

Cumulating all the requests received from China, Japan and Korea, there is a spectacular increase in the innovative potential of this region.

In this way, Asia has doubled its number of invention applications since 2005, and now accounts for about 43% of the total.

In terms of the number of trademark applications filed globally, China ranks first, with a numerical upward trend since 2008. In 2017, China registered more brand applications than the world's major powers cumulatively.

## 4. STATISTICS AND STUDIES REGARDING THE STAGE OF PATENT PROTECTION WORLDWIDE

4.1 Approaching the issue with the exemplification of concrete situations in top areas

Next, the focus will be on patent issues that give their owners an exclusive right to an invention, even when the issue is raised through the prism of a product / process, whose purpose is to come up with the initiative of a new solution, technical nature whose main objective is to solve problems.

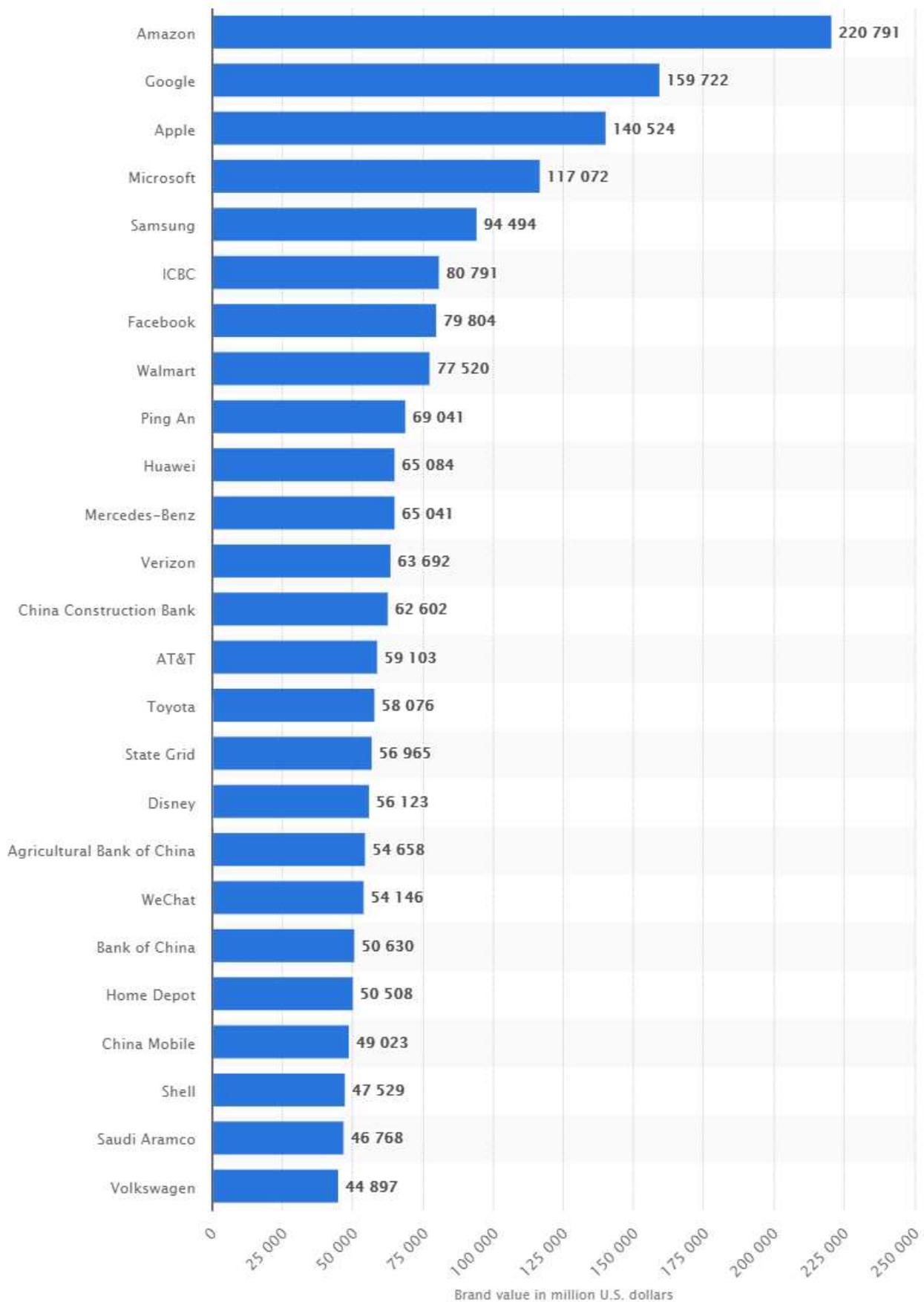

**Figure 10.** Ranking of brand value in millions of dollars [23]

Figure 10 shows the statistical situation of the most valuable brands registered in 2019.

From this figure it results that Amazon was the most valuable brand in the world since 2019, registering an estimated value of the brand of about 220 billion dollars. Google, the world's second-largest brand, had a $ 159 billion mark.

In third place was Apple with a registered value of 140 billion dollars.

### 4.2 Approaching the subject with exemplification in the medical field

The topic of patent statistics in the field of medicine was also examined. In 2018, more than 9,300 medical technology patents were granted.

Approximately 35% of French patents for diagnostic products filed in France, 24% were patents for implants and prostheses, 15% mainly for dialysis, infusion, resuscitation, syringes, catheters, 6% for dental equipment, 6% for hospital teams such as doctors. Beds, operating tables, wheelchairs, 5% patents for psychotherapy devices, 10% medical containers and others.

The United States was the largest target market for medical devices with almost 50% of the world market. These devices have as different goals as possible, as well as design and complexity. Therefore, the ability to develop the most effective patent strategy possible can be extremely valuable and equally complicated.

Regarding patent trends, industry registrations increased dramatically in 2015 (from around 22,400 in 2007 to around 34,400 in 2018), after which registrations decreased (to around 10,400 in 2018) [24].

The largest relative increase in registrations corresponds to the group of portable devices for which the registration of patents was 359% higher in 2018 than in 2007. The next significant increase was in the group of surgical devices, which increased by 181%. At the same time, the group of aids reached around 24,000 registrations in 2018.

With regard to the trends in patent applications in the auxiliary group, the situation is presented as follows [25]. Telemedicine teams dominated record growth in this group as their registrations increased 2.5-fold between 2007 and 2018. Telemedicine innovations can promote remote health management. This technology can be very useful in remote areas and / or offered to society at an accelerated pace. The global telemedicine market is estimated to be $ 21.56 billion in 2017 and $ 93.5 billion in 2026.

With regard to applicants for patents for medical devices by country of origin, the contribution of US applicants to US registrations in the medical device industry decreased slightly (from 67% in 2007 to 62%) according to [26]. % in 2018), even though the US records had over 33% in the same period. The contribution of Chinese applicants and applicants from South Korea has more than doubled, although contributions are still low.

Therefore, the medical device industry tends to incorporate new and different types of technologies. Patented devices are no longer simple devices that can be used, for example, in implants and in surgery. In contrast, portable devices dominate the industry. The medical device industry has a very high market value, but possibly high costs for regulatory approvals. Patent tracking data is used as an indicator of market movements and can be used to inform about these decisions.

### 5. FINAL CONCLUSIONS

One of the elements that makes a difference in the cultural, social and economic development of intellectual property.

Intellectual property protection offers strategic advantages and the ability to effectively develop and use new technologies.

Intellectual property is an indisputable feature of the market economy and at the same time a key feature of the world economy in global trade.

Today, the role of intellectual property is a major concern in both developed and developing countries. In this context, it is said that solutions are needed both for economic growth, development and competitiveness, and for increasing employment wealth.

Intellectual property rights include a variety of marketing tools or management strategies, which are monopoly rights that give the owner the exclusive right to use the property without the consent of the right person and prohibit use by a third party.

In today's economic environment, increasing the value of the company over time is the basic goal of every entrepreneur.

The company strives to leverage existing resources as efficiently as possible to produce and market products or services, with the primary objective of achieving a large market share while maintaining strong market share increase profits.

The technology sector is growing faster in business. Companies developing and marketing new technologies face patent protection issues. The patent process is not as fast as technological advances. However, this does not prevent technology companies from protecting their intellectual property or competitors who abuse their creations.

### 6. REFERENCES

1. Koh Winston, T.H., Poh Kam, Wong, *Competing at the Frontier: The Changing Role of Technology Policy in Singapore's Economic Strategy*, "Technological Forecasting and Social Change", Vol. 72, Issue 3, (2005).
2. Oprean, C., Țîțu, M., Bucur, V., "*The global management of the knowledge based organization*", Bucharest, Romania: AGIR, (2011).
3. Dragomirescu, H., *Economy, organizations, society in the information age*, vol I. Information economy, Publishing House ASE, Bucarest (2009).
4. Țîțu, M.A. et. al., The Place and the Role of the Intellectual Property Assets in the Knowledge Based Organization Context", Balkan Region Conference on Engineering and Business Education, Sibiu, Romania, (2015).
5. Bratianu, C., "*Knowledge Management. Fundamental Concepts*", Universitary Publishing House, Edition I, ISBN 978-606-28-0199-1, (2015).
6. Bratianu, C., "*Strategic Thinking*", Pro Universitaria Publishing House, ISBN: 9786062603991, (2015).
7. Durst, S., Gueldenberg, S., "The meaning of intangible assets: new insights into external company succession in SMEs", *Electronic Journal of Knowledge Management*, 7 (4), 437-446, (2012).
8. Tsai, Chih-Fong, et al., Intangible assets evaluation: The machine learning perspective", *Neurocomputing*, (2016).
9. Han, D. & Han, I., "*Prioritization and selection of intellectual capital measurement indicators using analytic hierarchy process for the mobile telecommunications industry*", Expert Systems with Applications, 26 (4), 519-527, (2004).
10. Slătineanu, L., "*Industrial property*", Iași, Romania: PERFORMANTICA, (2015).
11. Țîțu, M.A. et. al., "Quality Management of Intangible Assets in the Context of the Knowledge-Based Economy", *Proceedings of the 9th International Management Conference "Management and Innovation For Competitive*

Advantage", November 5th-6th, Bucharest, Romania, (2015);
12. Weltz, A.G., Fichtinger, M., Kerschbaum, F., „*Analiza de Status Quo pentru valorificarea proprietăţii intelectuale în zona Europei de Sud Est şi în context Global*", Viena: Institutul "Economica" de Cercetare Economică, (2013).
13. Sveiby, K.E., *The new organizational wealth: Managing & measuring knowledge-based assets*. San Francisco: Berrett-Koehler Publishers, Edvinsson and Malone (1997).
14. Edvinsson, L., M.S. Malone, , Intellectual capital: Realizing your company's true value by finding its hidden brainpower. New York: Haper Business, (1997).
15. ORDA, Romanian Copyright Office. Retrieved from https://www.orda.ro, (2017).
16. Săvescu, D., Budală A. *Proprietatea intelectuală în România şi unele ţări din UE*, ISBN 978-973-131-051-0, Braşov: Lux Libris, (2008).
17. https://en.wikipedia.org/wiki/European_Patent_Office, June 6, (2018).
18. Stuart J. H., Graham, H., Bronwyn, Dietmar Harhoff, D.C. Mowery, 2002 Post-Issue Patent "*Quality Control*": A Comparative Study of US Patent Re-Examinations and European Patent Oppositions, National Bureau of Economic Research 1050 Massachusetts Avenue, Cambridge, Ma 02138, February (2002).
19. https://osim.ro/despre-osim/statistici-publicate-in-2020/, accesed in may 2020.
20. https://www.epo.org/about-us/annual-reports-statistics/annual-report/2018/statistics/patent-applications.html, accesed in may 2020.
21. https://www.epo.org/about-us/annual-reports-statistics/annual-report/2018/statistics/patent-applications.html, accesed in may 2020.
22. https://www3.wipo.int/ipstats/ipsBarchartval, accesed in may 2020.
23. https://www.statista.com/statistics/264875/brand-value-of-the-25-most-valuable-brands/, accesed in may 2020.
24. https://www.ipwatchdog.com / 2019/05/08 / patented Trends-part-six-study medical-devices-industry-/ id = 108 972 /, accesed in may 2020.
25. https://www.ipwatchdog.com/2019/05/08/patent-trends-study-part-six-medical -devices-industry / id = 108 972 /, accesed in may 2020.
26. https://www.ipwatchdog.com/2019/05/08/patent-trends-study-part-six-medical- devices-industry / id = 108972 /, accesed in may 2020.